\documentstyle[preprint,aps,epsf]{revtex}
\begin{document}

\setlength{\parindent}{3em}
\setlength{\parskip}{1ex}
\newlength{\oldbaselineskip}

\draft
\preprint{\setlength{\baselineskip}{2.6ex}\hfil 
\vbox{\hbox{ \\}\hbox{TRI--PP--96--57} %numbered on Oct 10, 1996
%\hbox{September 24 1996}}}
\hbox{October 1996}}}

\title{T-Violation in $K^+ \rightarrow \mu^+ \nu \gamma$ Decay 
And Supersymmetry} 
 
\author{Guo-Hong Wu \footnote{\footnotesize gwu@alph02.triumf.ca
\vspace{-.15in}}
 and  John N. Ng \footnote{\footnotesize misery@triumf.ca}}

\address{TRIUMF Theory Group\\4004 Wesbrook Mall, Vancouver, B.C.,
Canada V6T 2A3}
                                                                               
\setlength{\baselineskip}{24pt}
\setlength{\oldbaselineskip}{\baselineskip}
\maketitle

\vskip -.3in
\begin{abstract}
%\addtolength{\baselineskip}{-.2\baselineskip}
Measurement of the transverse muon polarization
 $P^{\bot}_{\mu}$ in the $K^+ \rightarrow \mu^+\nu\gamma$ decay 
will be attempted for the first time at the ongoing KEK E246 experiment 
and also at a proposed BNL experiment.
 We provide a general analysis of how $P^{\bot}_{\mu}$ is sensitive
to the physical $CP$-violating phases in new physics induced
four-Fermi interactions,  
and then we calculate the dominant contributions to
$P^{\bot}_{\mu}$ from  
squark family mixings in generic supersymmetric models.
Estimates of the upper bounds on $P^{\bot}_{\mu}$ are also given.
  It is found that a supersymmetry-induced right-handed quark current
from $W$ boson exchange 
gives an upper limit on $P^{\bot}_{\mu}$ as large as a few per cent,
whereas with charged-Higgs-exchange induced pseudoscalar interaction,
$P^{\bot}_{\mu}$ is no larger than a few tenths of a per cent.
  Possible correlations between the muon polarization measurements
in $K^+ \rightarrow \mu^+\nu\gamma$ and $K^+ \rightarrow \pi^0\mu^+\nu$
decays are discussed,  and distinctive patterns of this correlation
from squark family-mixings and from the three-Higgs-doublet model 
are noted.
\end{abstract} 
%\vskip 0.2in
%\pacs{PACS numbers: 12.34}

%\setlength{\baselineskip}{\oldbaselineskip}
\newpage
 
\section{Introduction}
\label{sec:intro}

   The on-going KEK E246 experiment \cite{kuno} and a recently 
proposed BNL experiment \cite{adair}
are both devoted to testing $T$-violation to a high precision in 
the $K^+ \rightarrow \pi^0 \mu^+ \nu$ ($K^+_{\mu3}$) decay 
by measuring the transverse
muon polarization 
$P^{\bot(\pi)}_{\mu} = \bf{s_{\mu}}\cdot(\bf{p_{\pi}} \times \bf{p_{\mu}})
/|\bf{p_{\pi}} \times \bf{p_{\mu}}|$,
where $\bf{p_{\pi}}$ and $\bf{p_{\mu}}$ are the momenta of the pion
and muon in the kaon rest frame and $\bf{s_{\mu}}$ is the muon spin vector.
The combined previous measurements \cite{exp} at the BNL-AGS
constrained the muon polarization to be 
$P^{\bot(\pi)}_{\mu} = (-1.85 \pm 3.60) \times 10^{-3}$, and this 
puts an upper limit of 
$|P^{\bot(\pi)}_{\mu}| < 0.9 \%$ at the $95 \%$ confidence level.
 The standard model (SM) $CP$-violating contribution to 
 $P^{\bot}_{\mu}$ is vanishingly small \cite{GV},
and the final state interaction (FSI) effect is found to be of order
$10^{-6}$ \cite{zhit}. 
Therefore, if an effect is detected at the $10^{-3}$ level or $10^{-4}$
level which the KEK experiment and the proposed BNL experiment 
are respectively sensitive to, it  will be an unmistakable
signature for new physics. 
It has been estimated \cite{kmu3} that $P^{\bot(\pi)}_{\mu}$ can be 
as large as $\sim 10^{-3}$ in the three-Higgs-doublet model \cite{wei}.
More recently, we noted that \cite{wn} 
large squark-family-mixings in supersymmetry (SUSY) could contribute to
$P^{\bot(\pi)}_{\mu}$ at the level of $10^{-3}$, which is three orders of 
magnitude larger than that in the absence of squark-family-mixings 
\cite{CF}.

  The transverse muon polarization, 
denoted by $P^{\bot}_{\mu}$ and defined as above with $\bf{p_{\pi}}$
substituted by the photon momentum $\bf{p_{\gamma}}$,  will also be measured
  in the radiative decay mode $K^+ \rightarrow \mu^+ \nu \gamma$
($K_{\mu2\gamma}$) both at KEK \cite{kuno} and at BNL \cite{adair}.
 Here the FSI effect is expected to be  large \cite{mar},
  and it could be on the order of $10^{-3}$ \cite{geng}. 
Being electromagnetic in nature, this effect
can be accurately computed and subtracted out.
 The more interesting standard model $CP$-violating contribution 
to $P^{\bot}_{\mu}$ 
arising from the Kobayashi-Maskawa (KM) phase \cite{KM} is again negligible.
 The effect  of a $CP$-violating  tensor interaction  on the muon polarization
 was considered in Ref.~\cite{gengT}; 
 and the contribution  to $P^{\bot}_{\mu}$ 
 from an effective pseudoscalar four-Fermi operator
in the three-Higgs-doublet model (3HDM)
  has recently been discussed \cite{KLO}. 
In this work, we provide a more general analysis of 
the muon polarization in the $K_{\mu2\gamma}$ decay
by including the complete effective four-Fermi interactions induced
from spin-zero and spin-one boson exchange.
 Then we  concentrate on supersymmetric theories with large 
squark-generational-mixings where  dramatic enhancement effects 
due to the third family heavy quark masses could give rise
to a large $P^{\bot}_{\mu}$.  Details of this will be given in a later
section where SUSY effects are examined.
  
The outline of the paper is as follows. The framework
is laid out in section~\ref{sec:GF} 
for computing $P^{\bot}_{\mu}$ in $K_{\mu2\gamma}$ decay
in terms of general effective four-Fermi interactions.
In section~\ref{sec:SUSY}, we focus on the effects of large 
squark-family-mixings which are allowed in fairly general SUSY models.
Possible correlations of the muon polarization in 
$K_{\mu3}$ and $K_{\mu2\gamma}$ decays are then discussed,
 and an interesting  comparison with multi-Higgs type models is made.
 The conclusions are presented in section~\ref{sec:concl}.

\section{General Framework}
\label{sec:GF}

Consider the radiative $K^+_{\mu2}$ decay
\begin{eqnarray}
K^+ (p) & \rightarrow &   \gamma (q) \mu^+ (l) \nu (p_{\nu}),
\end{eqnarray}
where $p$, $q$, $l$ and $p_{\nu}$ denote the momenta of the 
kaon, photon, muon, and neutrino respectively.
 The SM amplitude for this decay consists of 
two separately gauge invariant pieces:
 the inner bremsstrahlung (IB) piece with the photon radiated off the
external muon or kaon line, and the structure-dependent (SD) piece 
for which the photon is emitted from the effective $K\mu\nu$ vertex via
some intermediate states. The total amplitude can be written 
as \cite{bry,eck},
\begin{eqnarray}
{\cal M}_{SM} & = & {\cal M}_{IB} + {\cal M}_{SD} \label{eq:ampl}\\
{\cal M}_{IB} & = &  - ie \frac{G_F}{\sqrt{2}} \sin \theta_c f_K m_{\mu}
\epsilon^*_{\alpha} K^{\alpha}    \label{eq:IB}  \\
{\cal M}_{SD} & = & ie \frac{G_F}{\sqrt{2}} \sin \theta_c  
\epsilon^*_{\alpha} L_{\beta} H^{\alpha\beta} ,   \label{eq:SD}
\end{eqnarray}
with
\begin{eqnarray}
K^{\alpha} & = & \overline{u}(p_{\nu})(1+\gamma_5) 
\left( \frac{p^{\alpha}}{p \cdot q} - \frac{2l^{\alpha} + \not\!q 
\gamma^{\alpha}} {2l \cdot q} \right) v(l) \\
L^{\alpha} & = & \overline{u}(p_{\nu})\gamma^{\alpha} (1-\gamma_5) v(l) \\
H^{\alpha\beta} & = & \frac{F_A}{m_K} (-g^{\alpha\beta} p \cdot q
 + p^{\alpha} q^{\beta} ) 
 + i \frac{F_V}{m_K} \epsilon^{\alpha\beta\mu\nu} q_{\mu} p_{\nu},
\label{eq:H}
\end{eqnarray}
where $G_F$ is the Fermi constant, $\sin \theta_c=0.22$ is the Cabibbo
mixing, $m_{\mu}$ and $m_K$ are the masses of the muon and the kaon,
$\epsilon$ is the photon polarization vector,
$f_K$ is the well determined kaon decay constant, 
and $F_A$ and $F_V$ are the axial-vector
and vector form factors associated with the radiative decay. 
The kaon decay constant and the two form factors are 
 defined by
\begin{eqnarray}
\langle 0 |\overline{s} \gamma^{\mu} \gamma_5 u |K^+(p) \rangle  & = &
 - if_K p^{\mu}      \label{eq:meA} \\
\int dx e^{iqx} \langle 0 | T(J^{\mu}_{em}(x) 
\overline{s} \gamma^{\nu} \gamma_5 u(0))| K^+(p) \rangle  & = &
 - f_K \left( g^{\mu\nu} + \frac{p^{\mu}(p-q)^{\nu}}{p \cdot q} \right)
 \nonumber \\
&&
+\frac{F_A}{m_K}(g^{\mu\nu}p \cdot q -p^{\mu} q^{\nu})  \label{eq:meAv} \\
\int dx e^{iqx} \langle 0 | T(J^{\mu}_{em}(x)
\overline{s} \gamma^{\nu} u(0))| K^+(p) \rangle  & = & 
i\frac{F_V}{m_K} \epsilon^{\mu\nu\alpha\beta} q_{\alpha} p_{\beta} ,
\label{eq:meVv}
\end{eqnarray}
where $f_K=160\;\mbox{MeV}$,
$F_A$ and $F_V$ are functions of $(p-q)^2$,
 $J^{\mu}_{em}$ is the electromagnetic current, and 
$\epsilon_{0123}=1$.

 Ideally one would like to be able to extract separately $F_V$ and $F_A$
from experimental data. However, the accuracy of current data does not permit
us to do so.
On the other hand, 
various models \cite{models} have been used to calculate the form factors.
In chiral perturbation theory at the one-loop level,
$F_V$ and $F_A$ are found to be real and are given by \cite{eck} 
\begin{equation}
F_V =  -0.0945~~~~~~F_A = -0.0425 .
\end{equation}
The momentum dependence of $F_V$ and $F_A$ shows up only at two loops
in chiral  perturbation theory. 
Furthermore in the SM, the KM phase will enter the form factors
at the two-loop level and hence can be ignored.
The above estimate will be used in our analysis of the 
muon polarization.

 As will be shown later, contributions to the 
$K^+ \rightarrow \mu^+ \nu \gamma$ decay 
  from physics beyond the SM 
can be parameterized by three dimensionless quantities,
$\delta_{IB}$, $\delta_A$ and $\delta_V$, to be associated with
$f_K$, $F_A$, and $F_V$ respectively.
The new amplitude is obtained from Eqs.~(\ref{eq:ampl})-(\ref{eq:H})
by the following  replacements:
\begin{eqnarray}
f_K & \rightarrow & f_K^{\prime} \equiv f_K(1 + \delta_{IB}) 
\label{eq:defdelIB}\\
F_A & \rightarrow & F_A^{\prime} \equiv F_A(1 + \delta_A )   
\label{eq:defdelA}  \\
F_V & \rightarrow & F_V^{\prime} \equiv F_V(1 +\delta_V)
\label{eq:defdelV} 
\end{eqnarray}
The three $\delta$ parameters are in general complex, and could contribute 
to the $T$-odd transverse muon polarization $P^{\bot}_{\mu}$.

   The transverse polarization of the muon in 
$K^+ \rightarrow \mu^+ \nu \gamma$ decay is defined as
\begin{eqnarray}
P^{\bot}_{\mu} & = & \frac{\bf{s_{\mu}} \cdot (\bf{p_{\gamma}} 
\times \bf{p_{\mu}})}
   {|\bf{p_{\gamma}} \times \bf{p_{\mu}}|},
\label{eq:defpol}
\end{eqnarray}
where $\bf{s_{\mu}}$ is the spin vector of the muon, and 
$\bf{p_{\gamma}}$ and
$\bf{p_{\mu}}$ are the three-momenta of the photon and muon.
A non-zero $P^{\bot}_{\mu}$ arises from the interference between the 
${\cal M}_{IB}$ and ${\cal M}_{SD}$ amplitudes.
After a general kinematic analysis, we can express $P^{\bot}_{\mu}$
in the $K^+$ rest frame as the sum of an $IB$-$F_V$
interference piece and an $IB$-$F_A$ interference piece. Explicitly,
\begin{eqnarray}
P^{\bot}_{\mu}(x,y)& = & P^{\bot}_{IB-V}(x,y) + 
                 P^{\bot}_{IB-A}(x,y) \label{eq:pol} \\
P^{\bot}_{IB-V}(x,y) & = & \sigma_V(x,y)Im[ (1+\delta_{IB})(1+\delta_V^*)] 
 \label{eq:polV} \\
P^{\bot}_{IB-A}(x,y) & = & \sigma_A(x,y) Im[ (1+\delta_{IB})(1+\delta_A^*)]  
 , \label{eq:polA}  
\end{eqnarray}
where $x=2p \cdot q/p^2=2E_{\gamma}/m_K$ and $y=2p \cdot l/p^2=2E_{\mu}/m_K$ 
are the normalized energies of the photon and the muon respectively.
 The functions $\sigma_V(x,y)$ and $\sigma_A(x,y)$ are given by 
\begin{eqnarray}
\sigma_V(x,y) & = & - 2 \sqrt{r_{\mu}} \frac{f_K}{m_K} F_V f_V(x,y) 
 \frac{\sqrt{(1-y+r_{\mu})((1-x)(x+y-1)-r_{\mu})}}{\rho(x,y)}
 \\
\sigma_A(x,y) & = & - 2 \sqrt{r_{\mu}} \frac{f_K}{m_K} F_A f_A(x,y) 
 \frac{\sqrt{(1-y+r_{\mu})((1-x)(x+y-1)-r_{\mu})}}{\rho(x,y)},
\end{eqnarray}
where $r_{\mu}=m^2_{\mu}/m^2_K$, $f_V(x,y)$ and $f_A(x,y)$ are to be 
defined later, and 
$\rho(x,y) \propto \frac{d\Gamma (K^+ \rightarrow \mu^+ \nu \gamma)}{dxdy}$
 is the normalized Dalitz density consisting of  
the IB piece $\rho_{IB}(x,y)$, 
the SD piece $\rho_{SD}(x,y)$, and the
interference term $\rho_{INT}(x,y)$,
\begin{eqnarray}
\rho(x,y) & = & \rho_{IB}(x,y) +  \rho_{SD}(x,y) + \rho_{INT}(x,y) 
\label{eq:rhoxy}  \\
\rho_{IB}(x,y)  & = & 
 2 r_{\mu} \frac{f^2_K}{m^2_K} |1+\delta_{IB}|^2 f_{IB}(x,y) \nonumber \\
\rho_{SD}(x,y) & = & \frac{1}{2} (|F_V^{\prime} +F_A^{\prime}|^2 f_{SD^+}(x,y)
 + |F_V^{\prime} -F_A^{\prime}|^2 f_{SD^-}(x,y)) \nonumber \\
\rho_{INT}(x,y) & = &  2 r_{\mu} \frac{f_K}{m_K} 
Re[(1+\delta_{IB}^*)((F_V^{\prime} +F_A^{\prime}) f_{INT^+}(x,y)
    + (F_V^{\prime} -F_A^{\prime}) f_{INT^-}(x,y))],  \nonumber
\end{eqnarray}
with
\begin{eqnarray}
f_{IB}(x,y) & = & \left( \frac{1-y+r_{\mu}}{x^2(x+y-1-r_{\mu})} \right)
   \left( x^2+ 2(1-x)(1-r_{\mu}) - \frac{2xr_{\mu}(1-r_{\mu})}{x+y-1-r_{\mu}} 
   \right)
\nonumber \\
f_{SD^+}(x,y) & = & (x+y-1-r_{\mu})((x+y-1)(1-x)-r_{\mu})
\nonumber \\
f_{SD^-}(x,y) & = & (1-y+r_{\mu})((1-x)(1-y)+r_{\mu})
\nonumber \\
f_{INT^+}(x,y) & = & \left( \frac{1-y+r_{\mu}}{x(x+y-1-r_{\mu})} \right) 
 ((1-x)(1-x-y)+r_{\mu}) \nonumber \\
 f_{INT^-}(x,y)  & = & \left( \frac{1-y+r_{\mu}}{x(x+y-1-r_{\mu})} \right)
 (x^2-(1-x)(1-x-y)-r_{\mu})  \nonumber \\
f_V(x,y) & = & \frac{2-x-y}{x+y-1-r_{\mu}}
\nonumber \\
f_A(x,y) & = & \frac{(2-x)(x+y)-2(1+r_{\mu})}{x(x+y-1-r_{\mu})}.
\end{eqnarray}

 The next step in the analysis is to compute the $\delta$ parameters in 
Eqs.~(\ref{eq:defdelIB})-(\ref{eq:defdelV}) in terms of parameters
describing the new physics.
It is reasonable to assume that at the energies we are considering, 
new physics can be described by general four-Fermi 
operators of the form (neglecting possible tensor interactions),  
\begin{eqnarray}
{\cal L} & = & - \frac{G_F}{\sqrt{2}} \sin \theta_c 
        \overline{s} \gamma_{\alpha} (1 - \gamma_5) u 
        \overline{\nu} \gamma^{\alpha} (1- \gamma_5) \mu
 \nonumber \\
   & &     +G_S \overline{s} u \overline{\nu} (1 + \gamma_5) \mu 
        + G_P \overline{s} \gamma_5 u  \overline{\nu} (1 + \gamma_5) \mu
 \nonumber \\
   & &       + G_V \overline{s} \gamma_{\alpha}  u 
         \overline{\nu} \gamma^{\alpha} (1- \gamma_5) \mu
            + G_A \overline{s} \gamma_{\alpha} \gamma_5 u
         \overline{\nu} \gamma^{\alpha} (1- \gamma_5) \mu
 \nonumber \\
  & & + h.c. , \label{eq:interaction}
\end{eqnarray}
where $G_S$, $G_P$, $G_V$ and $G_A$ 
parameterize the non-standard model interactions due to scalar,
pseudoscalar, vector and axial-vector boson exchange respectively.
In some models, effective right-handed neutrino-muon current
can be constructed. Since the  SM leptonic charged-current is left-handed,
its interference with such a right-handed current 
will be suppressed by the neutrino mass and will have a negligible
contribution to $P^{\bot}_{\mu}$. 
Therefore only  left-handed neutrinos need to be considered.

    The contributions of the $G_S$ and $G_P$ operators are
recently discussed  by Kobayashi {\it et. al.} \cite{KLO} 
in the context of the three-Higgs-doublet model, where tree-level
charged Higgs exchange is sufficient to give rise to $T$ violation.
The relevant hadronic matrix elements involving the scalar current
can be related to $f_K$ via \cite{KLO}
\begin{eqnarray}
\langle 0 |\overline{s} \gamma_5 u |K^+(p) \rangle  & = &
i \frac{f_K m_K^2}{m_s + m_u}       \label{eq:meP} \\ 
\int dx e^{iqx} \langle 0 | T(J^{\mu}_{em}(x) 
\overline{s} \gamma_5 u(0))| K^+(p) \rangle  & = &
  \frac{p^{\mu}}{p \cdot q} \frac{f_K m_K^2}{m_s + m_u}  \label{eq:mePv} \\ 
\int dx e^{iqx} \langle 0 | T(J^{\mu}_{em}(x)
\overline{s} u(0))| K^+(p) \rangle  & = & 0= 
\langle 0 |\overline{s} u |K^+(p) \rangle .
\end{eqnarray}
 It is immediately seen that the effective scalar interaction 
(the $G_S$ operator)
does not affect the $K_{\mu2\gamma}$ decay rate as their hadronic 
matrix elements vanish by parity.
On the other hand from Eqs.~(\ref{eq:meP}) and (\ref{eq:mePv}), 
the contribution of the  $G_P$ operator
to the decay amplitude is seen to be non-vanishing
and it has the same structure as ${\cal M}_{IB}$ of Eq.~(\ref{eq:IB}), 
\begin{eqnarray}
{\cal M}_P & = & - ie G_P \frac{f_K m_K^2}{m_s + m_u}
 \epsilon^*_{\alpha} K^{\alpha}.
\end{eqnarray}
This amounts to a contribution to the parameter $\delta_{IB}$ only,
\begin{eqnarray}
\delta_{IB}|_{G_P} & = & \frac{\sqrt{2}G_P}{G_F\sin\theta_c}
      \frac{m_K^2}{(m_s+m_u)m_{\mu}}, 
\label{eq:delIBGP}
\end{eqnarray} 
and the other two parameters, $\delta_V$ and $\delta_A$, remain zero.

  The effective  vector and axial-vector four-Fermi interactions
give rise to corrections to the $V-A$ structure of the SM quark current.
They can be analyzed by using 
the corresponding hadronic matrix elements given by
Eqs.~(\ref{eq:meA})-(\ref{eq:meVv}).
   The  $G_V$ operator contributes only to $\delta_V$, 
\begin{eqnarray}
\delta_V|_{G_V} & = & - \frac{\sqrt{2}G_V}{G_F\sin \theta_c} .
\label{eq:delVGV}
\end{eqnarray}
On the other hand, the axial-vector $G_A$ operator contributes to both  
$\delta_{IB}$ and $\delta_A$,
\begin{equation}
\delta_{IB}|_{G_A}  = \delta_A|_{G_A} =  
\frac{\sqrt{2}G_A}{G_F \sin \theta_c}.
\label{eq:delAGA}
\end{equation}

  Summing up Eqs.~(\ref{eq:delIBGP})-(\ref{eq:delAGA}), 
we obtain the following contributions to the $\delta$ parameters
 from the effective four-Fermi interactions
\begin{eqnarray}
\delta_{IB} & = &  \frac{\sqrt{2}G_P}{G_F\sin\theta_c}
      \frac{m_K^2}{(m_s+m_u)m_{\mu}}
     + \frac{\sqrt{2}G_A}{G_F \sin \theta_c}     \label{eq:deltaIB}  \\
\delta_V & = & - \frac{\sqrt{2}G_V}{G_F\sin \theta_c}   \label{eq:deltaV} \\
\delta_A & = &  \frac{\sqrt{2}G_A}{G_F \sin \theta_c}.   \label{eq:deltaA}
\end{eqnarray}

 As can be seen from Eqs.~(\ref{eq:pol})-(\ref{eq:polA}) and
Eqs.~(\ref{eq:deltaIB})-(\ref{eq:deltaA}),
 $P^{\bot}_{\mu}$ in the $K^+ \rightarrow \mu^+ \nu \gamma$ decay 
could receive contributions from the $G_P$, $G_V$ and $G_A$ effective 
interactions of Eq.~(\ref{eq:interaction}). 
By comparison, $P^{\bot(\pi)}_{\mu}$ in the
$K^+ \rightarrow \pi^0 \mu^+ \nu$ decay is only sensitive to 
the $G_S$ interaction \cite{cheng}.
These two decay modes are therefore  complementary in searching for 
new physics effects.
 As one would expect, if physics beyond the SM has only left-handed 
quark current, it  will not contribute to $P^{\bot}_{\mu}$.
This can be seen explicitly from
Eqs.~(\ref{eq:deltaIB})-(\ref{eq:deltaA}).
For a left-handed quark current ($G_V=-G_A$ and $G_P=G_S=0$),
$\delta_{IB}=\delta_A=\delta_V$, and therefore no relative phase exists
 between ${\cal M}_{IB}$ and ${\cal M}_{SD}$.
However, a right-handed quark current as in left-right symmetric models
will in general have an effect on $P^{\bot}_{\mu}$, 
 and the size of the muon polarization would be model dependent.
Signatures of the effective four-Fermi interactions of 
Eq.~(\ref{eq:interaction}) for the muon polarization in the
$K_{\mu2\gamma}$ decay  are summarized in Table~\ref{tab:signatures}.

\begin{figure}[p]
\epsfsize=100pt \epsfbox[80 150 800 710]{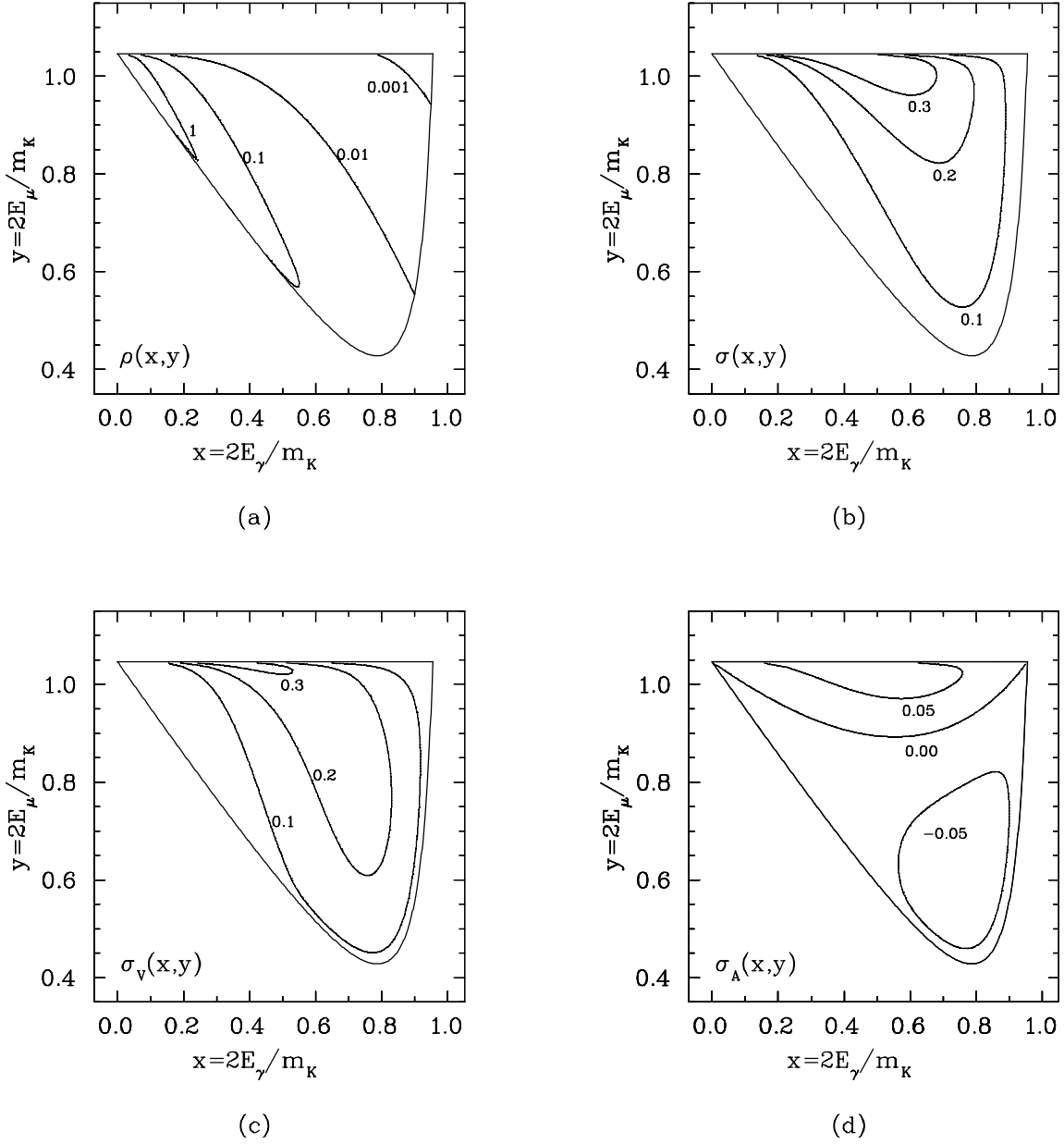}
\vspace{0pt} 
\caption{ Contour plots of the normalized Dalitz density $\rho(x,y)$
 (a), and the transverse muon polarization functions
  $\sigma(x,y)$ (b), $\sigma_V(x,y)$ (c), and $\sigma_A(x,y)$ (d)
 in the $K^+ \rightarrow \mu^+ \nu \gamma$ decay.}
\label{fig:conplots}
\end{figure}

   The success of the SM dictates that the magnitudes of the 
$\delta$ parameters of Eqs.~(\ref{eq:deltaIB})-(\ref{eq:deltaA}) 
are much smaller than one.
   We can therefore simplify the formula for $P^{\bot}_{\mu}$
 by neglecting terms quadratic in the $\delta$'s and 
 by considering only the relevant $G_P$ and $G_R$ interactions given by 
\begin{eqnarray}
\Delta {\cal L} & = & 
         G_P \overline{s} \gamma_5 u  \overline{\nu} (1 + \gamma_5) \mu
        + G_R \overline{s} \gamma_{\alpha} (1+ \gamma_5) u 
         \overline{\nu} \gamma^{\alpha} (1- \gamma_5) \mu
 + h.c. .  \label{eq:deltal} 
\end{eqnarray}
 Combining Eqs.~(\ref{eq:pol})-(\ref{eq:polA}) and  
(\ref{eq:deltaIB})-(\ref{eq:deltaA}),
   the muon transverse polarization can be rewritten as  a sum of 
the $G_P$ interaction contribution $P^{\bot}_{\mu,P}(x,y)$ and 
the $G_R$ interaction contribution $P^{\bot}_{\mu,R}(x,y)$, 
\begin{eqnarray}
P^{\bot}_{\mu}(x,y) & = &  P^{\bot}_{\mu,P}(x,y) + P^{\bot}_{\mu,R}(x,y)
\label{eq:polS} \\ 
P^{\bot}_{\mu,P}(x,y) & =  &  \sigma(x,y) Im\Delta_{P}
\label{eq:polP} \\
 P^{\bot}_{\mu,R}(x,y) & = &  2 \sigma_V(x,y)Im\Delta_R ,      
\label{eq:polR}
\end{eqnarray}
where
\begin{eqnarray}
\sigma(x,y) & = & \sigma_V(x,y) +\sigma_A(x,y) \\ 
\Delta_P & \equiv & \frac{\sqrt{2}G_P}{G_F\sin\theta_c}  
      \frac{m_K^2}{(m_s+m_u)m_{\mu}}  \label{eq:delP} \\
\Delta_R & \equiv & \frac{\sqrt{2}G_R}{G_F\sin\theta_c} .  \label{eq:delR}
\end{eqnarray}            
   The contour plots of $\rho(x,y)$, $\sigma(x,y)$,  
$\sigma_V(x,y)$, and $\sigma_A(x,y)$ are given in Fig.~\ref{fig:conplots}.
The infrared sensitivity of the $K_{\mu2\gamma}$ decay rate manifests
itself in the Dalitz plot $\rho(x,y)$ in the soft photon (i.e. small $x$) 
region.

   Due to the experimental necessity of cutting out low energy photons, it is 
more useful to define the quantity $\overline{P^{\bot}_{\mu}}$
by 
\begin{eqnarray}
\overline{P^{\bot}_{\mu}} & \equiv & \frac{
\int_S dx dy \rho(x,y) P^{\bot}_{\mu} (x,y) }
{\int_S dx dy \rho(x,y) }, \label{eq:polav}
\end{eqnarray} 
which is the average of $P^{\bot}_{\mu} (x,y)$ over a region of phase space
$S$. 
 In this definition, the numerator measures the difference in number 
in the region $S$ between muons pointing their spin vector along the direction
$\bf{p_{\gamma}} \times \bf{p_{\mu}}$ and those along the opposite 
direction, and 
the denominator is a measure of the total number of muons in the region
$S$.
In terms of the effective four-Fermi interactions, the average muon
polarization is given by
\begin{eqnarray}
\overline{P^{\bot}_{\mu}} & = & \overline{P^{\bot}_{\mu,P}} 
 + \overline{P^{\bot}_{\mu,R}} = 
 \overline{\sigma} Im\Delta_{P}
 + 2  \overline{\sigma_V} Im\Delta_R .      \label{eq:polave}
\end{eqnarray}

\begin{figure}[p]
\epsfsize=60pt \epsfbox[80 200 800 710]{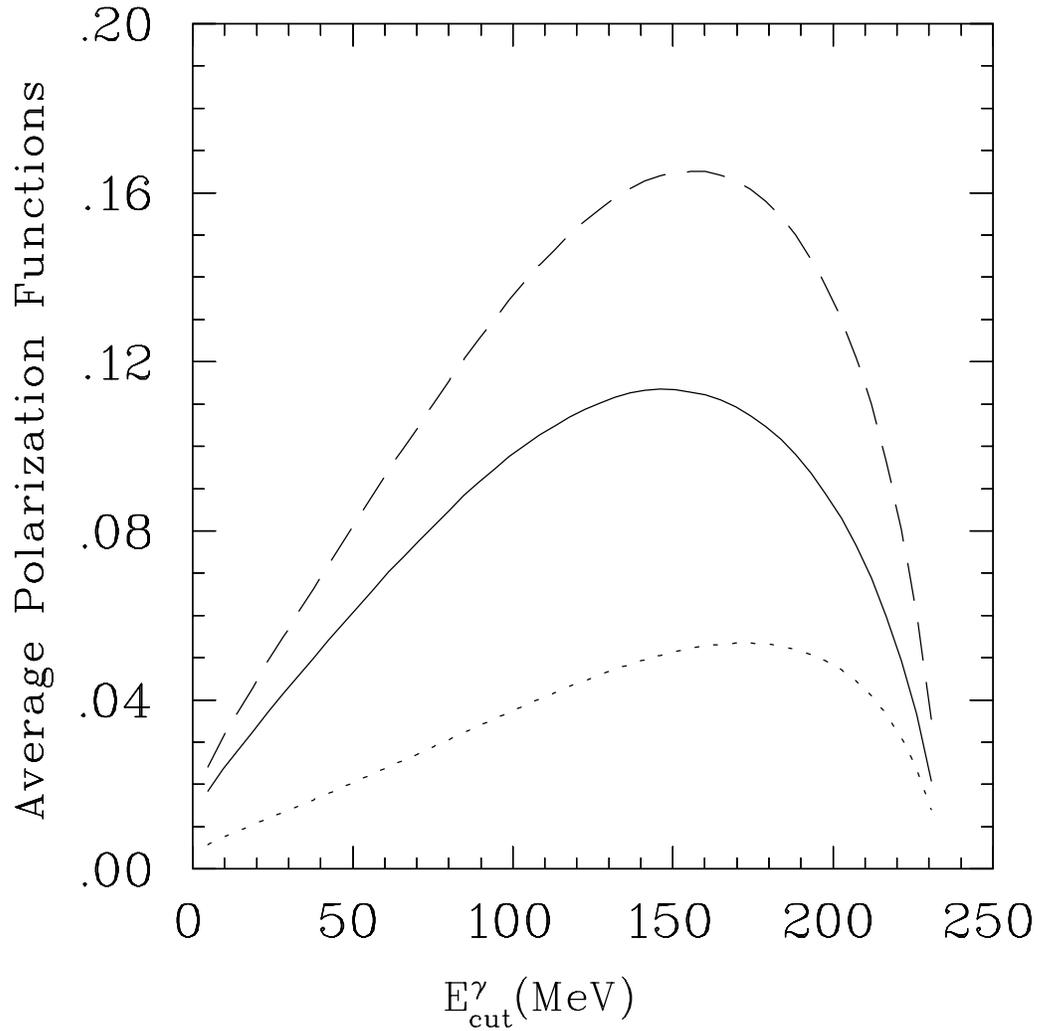}
\vspace{0pt} 
\caption{ The averaged transverse muon polarization functions vs. 
the soft photon energy cut $E_{cut}^{\gamma}$ for the
$K^+ \rightarrow \mu^+ \nu \gamma$ process:
$\overline{\sigma}$ (solid line),
$\overline{\sigma_V}$ (long-dashed line), and 
$-\overline{\sigma_A}$ (short-dashed line).}
\label{fig:avepol}
\end{figure}

  Plots of the averaged $\overline{\sigma}$,
$\overline{\sigma_V}$ and $-\overline{\sigma_A}$
as a function of the energy cut on soft photons are given in 
Fig.~\ref{fig:avepol}.
The typical size of $\overline{\sigma}$ and
$\overline{\sigma_V}$ for the $K^+_{\mu2\gamma}$ decay
can be seen to be of order $0.1$,
and it is a kinematic measure of the relative strength of the 
${\cal M}_{IB}$-${\cal M}_{SD}$ interference and the $IB$ plus $SD$ 
contributions to the partial width. The corresponding values
for the radiative decays of 
$\pi^+ \rightarrow \mu^+ \nu \gamma$ ($\pi_{\mu2\gamma}$) and 
$K^+ \rightarrow e^+ \nu \gamma$ ($K_{e2\gamma}$)
are expected to be roughly two orders of magnitude smaller 
than for $K_{\mu2\gamma}$ decay because of the dominance of the $IB$ and
$SD$ contributions respectively. And this makes it difficult to measure
the transverse lepton polarization in the $\pi_{\mu2\gamma}$ 
and $K_{e2\gamma}$ modes.

\section{SUSY Effects}
\label{sec:SUSY}

  In the minimal supersymmetric standard model (MSSM, by this we mean 
with minimal particle content and with $R$-parity conservation), 
$T$-violation in $K^+ \rightarrow \mu^+ \nu\gamma$
decay arises from the interference between the SM tree amplitude and 
the one-loop MSSM amplitude.
 Naively, this would be suppressed
at least by  $\alpha_s/\pi$ ($\alpha_s$ is the QCD coupling)
 relative to the tree level interference effect 
as in the three-Higgs-doublet model \cite{KLO}, 
and would be too small to be seen. 
 However, this would not be the case when the squark family mixings
are taken into account. Then $T$-violation in $K_{\mu2\gamma}$ decay
could be sensitive to the top and bottom quarks of the third family, and
large enhancement effects due to the heavy quark masses could 
appear \cite{wn}.  This scenario will be the focus of our attention 
in the discussion of the SUSY contributions to the muon polarization. 

  The notion of squark family mixings comes 
from the general assumption that the mass matrices of the
quarks and squarks are diagonalized by different unitary transformations
in generation space \cite{fcnc}. 
In principle, these mixing matrices could all be determined in specific
models. In the lack of a generally accepted SUSY flavor model, we adopt a
model-independent approach and refer interested readers to the literature
for discussions of specific models \cite{umix,uhall}.
The relative rotations in generation space 
between the $\tilde{u}_L$, $\tilde{u}_R$,
$\tilde{d}_L$, and $\tilde{d}_R$ squarks and their corresponding 
quark partners are denoted by $V^{U_L}$, $V^{U_R}$, $V^{D_L}$, 
and $V^{D_R}$ respectively.
These matrices appear in the quark-squark-gluino couplings
which lead to new contributions to flavor-changing neutral current processes
(FCNC) \cite{fcnc}. Meanwhile, these generational mixing matrices 
give rise to the interesting possibility that the heavy fermions of 
the third family may play an important role in low energy processes,
including the neutron electric dipole moment \cite{hall}.

 As FCNC processes occur only at loop level in both the SM and the MSSM,
severe constraints on the squark family mixings can be derived  
\cite{fcnc} based on available experimental data. 
For charged-current (CC) processes on the other hand,
the SM contributions often appear at tree level
whereas effects of squark family mixings arise only 
at loop-level,
 and current data in the hadronic sector are not precise enough to 
put useful bounds on the squark mixing matrices.
The FCNC constraints 
 can be written in the form of upper limits either on
 product of different $V^D$'s, as from $K\overline{K}$ and $B\overline{B}$
systems and from $b \rightarrow s \gamma$, or on product of different 
$V^U$'s, as from $D\overline{D}$ mixing.
 Without assumptions on or model preference for these mixing matrices,
the individual matrix element can however still be of order one. 

   Charged-current processes involve the product of 
$V^U$ and $V^D$, and the size of both mixing matrix elements can be of 
order one without violating the FCNC bounds.
Because of the large top Yukawa coupling, sizable squark mixings 
with the third generation could then lead to large enhancement effects 
in low energy charged-current processes.
Although this loop-level enhancement may not have significant effects on 
$CP$-conserving, tree-level CC processes, it could have 
dramatic  consequences for $CP$-violating  CC processes for which 
the standard model effects are negligible.
 This possibility has recently been discussed in the context of
$T$ violation in charged meson 
semileptonic decays \cite{wn}.  In this work, we would like to 
extend the analysis to the transverse muon polarization in
  the radiative $K_{\mu2}$ decay.

 The physical phases that are relevant for the transverse muon
polarization could come from both
the squark mixing matrices and other soft SUSY breaking
operators including the $A$ terms and the gaugino mass terms. 
For the sake of simplicity and a clearer 
illustration of the underlying physics, we concentrate on the 
phases in the squark mixing matrices.
 Mass-insertion approximation will be used for the 
$\tilde{t}_L$--$\tilde{t}_R$ and
$\tilde{b}_L$--$\tilde{b}_R$  mixings,
and $m_{\tilde{t}_L}=m_{\tilde{t}_R}=m_{\tilde{t}}$ and 
$m_{\tilde{b}_L}=m_{\tilde{b}_R}=m_{\tilde{b}}$  will be assumed 
for the mass parameters of the left and right top and bottom squarks.
With large generational squark mixings, 
 the dominating contributions to $P^{\bot}_{\mu}$ are expected to come from 
the $G_P$ and $G_R$ four-Fermi operators induced by 
 the $\tilde{t}$-$\tilde{b}$-$\tilde{g}$
loop diagrams with $W$ boson and charged Higgs boson exchange.
The muon polarization will then be directly proportional 
to $|V^{U_{L,R}}_{31} {V^{D_{L,R}}_{32}}^*|$.

\subsection{$W$ Exchange}

\begin{figure}[p]
\epsfsize=10pt \epsfbox[80 150 500 730]{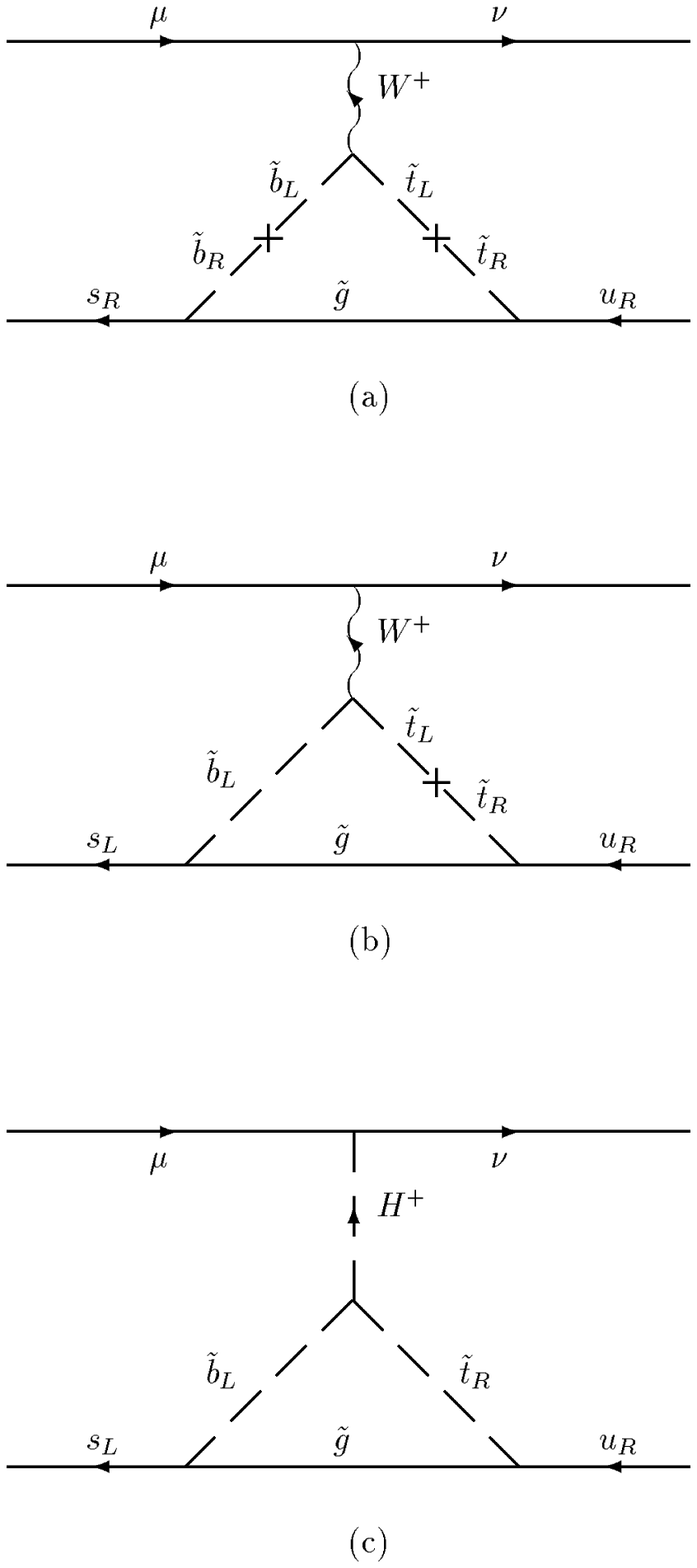}
\vspace{0pt} 
\caption{ Supersymmetry diagrams for
(a) the $W$-exchange induced effective $G_R$ interaction (see
 Eq.~(\ref{eq:L1})),
(b) the $W$-exchange induced effective $G_P$ interaction (see
 Eqs.~(\ref{eq:W}) and (\ref{eq:L2S})), and
(c)  the $H^+$-exchange induced effective $G_P$ interaction (see
 Eq.~(\ref{eq:LH}).}
\label{fig:diagrams} 
\end{figure}

\begin{figure}[p]
\epsfsize=100pt \epsfbox[80 200 800 710]{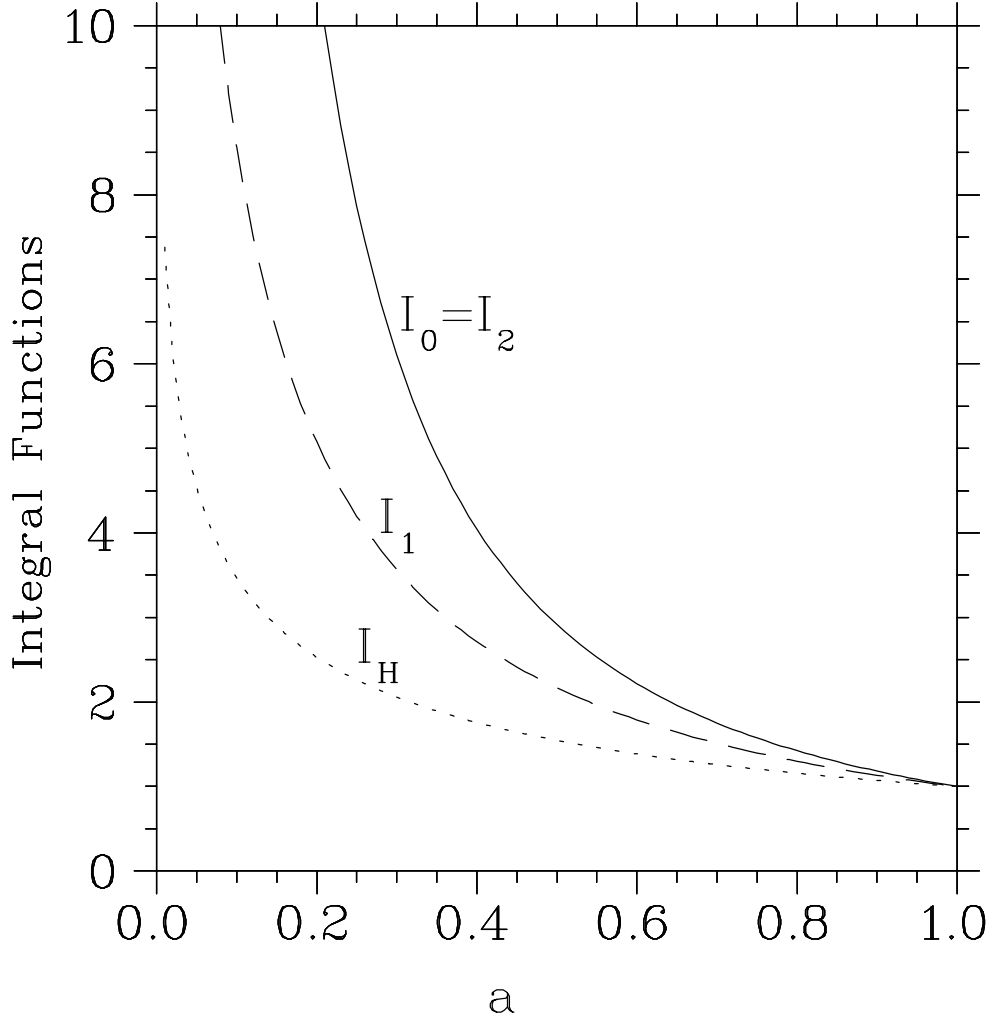}
\vspace{0pt}
\caption{ The integral functions vs. the parameter $a=
\frac{m_{\tilde{t}}^2}{m_{\tilde{g}}^2}=
\frac{m_{\tilde{b}}^2}{m_{\tilde{g}}^2}$ 
for $I_0$ (Eq.~(\ref{eq:I0})) and $I_2$ (Eq.~(\ref{eq:I2})) (solid line),
$I_1$ (Eq.~(\ref{eq:I1})) (long-dashed line), 
and $I_H$ (Eq.~(\ref{eq:IH})) (short-dashed line).}
\label{fig:4I}
\end{figure}

 An effective right-handed current interaction 
$W^{\mu} \overline{s_R} \gamma_{\mu} u_R$ 
can be generated by the diagram with $\tilde{g}$-$\tilde{t}$-$\tilde{b}$ 
sparticles  in the loop and with $\tilde{t}_L$--$\tilde{t}_R$ and
$\tilde{b}_L$--$\tilde{b}_R$ mass insertions 
(see Fig.~\ref{fig:diagrams}(a)). 
 This gives rise to an effective $G_R$ interaction,
\begin{eqnarray}
{\cal L}_1 & = &
 - \frac{4G_F}{\sqrt{2}} C_0 
(\overline{s}_R \gamma^{\alpha} u_R)(\overline{\nu}_L \gamma_{\alpha}\mu_L)
 + h.c. \label{eq:L1}
\end{eqnarray}
with
\begin{eqnarray}
C_0 & = & \frac{\alpha_s}{36 \pi} I_0 
         \frac{m_tm_b(A_t-\mu\cot\beta)(A_b-\mu\tan\beta)}
              {m_{\tilde{g}}^4}
 {V^{SKM}_{33}}^*V^{U_R}_{31}{V^{D_R}_{32}}^*  ,      
 \label{eq:C0}
\end{eqnarray}
where $\alpha_s\simeq 0.1$ is the QCD coupling 
evaluated at the mass scale of the sparticles in the loop,
$A_t$ and $A_b$ are the soft SUSY breaking $A$ terms for the top 
and bottom squarks,
$\mu$ denotes the two Higgs superfields mixing parameter, $\tan \beta$ is
the ratio of the two Higgs VEVs, $m_{\tilde{g}}$ is the mass of the  gluino, 
$V^{SKM}_{ij}$ is the super KM matrix associated with the 
$W$-squark coupling $W^+ {\tilde{u_i}_L}^* \tilde{d_j}_L$, and 
 the integral function $I_0$  is given by
\begin{eqnarray}
I_0 & = &  \int_0^1 dz_1 \int_0^{1-z_1} dz_2 \frac{24z_1z_2}{[
\frac{m_{\tilde{t}}^2}{m_{\tilde{g}}^2}z_1+
\frac{m_{\tilde{b}}^2}{m_{\tilde{g}}^2}z_2+
(1-z_1-z_2)]^2}.  \label{eq:I0} 
\end{eqnarray}
Note that $I_0=1$ for 
$\frac{m_{\tilde{t}}}{m_{\tilde{g}}}=\frac{m_{\tilde{b}}}{m_{\tilde{g}}}=1$,
but it increases rapidly to $\sim 8$ as the squark-to-gluino mass ratios 
decrease to
$\frac{m_{\tilde{t}}}{m_{\tilde{g}}}=\frac{m_{\tilde{b}}}{m_{\tilde{g}}}=
\frac{1}{2}$.
For the case of $m_{\tilde{t}}=m_{\tilde{b}}$, the variation of $I_0$ 
with the mass ratio is plotted in Fig.~\ref{fig:4I}. 

 From Eqs.~(\ref{eq:L1})-(\ref{eq:C0}) and Eq.~(\ref{eq:delR}), 
the right-handed current contribution to $\Delta_R$ is found to be
\begin{eqnarray}
\Delta_R|_{{\cal L}_1} 
  & = & -  \frac{\alpha_s}{36 \pi} I_0
         \frac{m_tm_b(A_t-\mu\cot\beta)(A_b-\mu\tan\beta)}
              {m_{\tilde{g}}^4} \times
 \frac{[{V^{SKM}_{33}}^*V^{U_R}_{31}{V^{D_R}_{32}}^*]}{\sin \theta_c}.
 \label{eq:deltaR}
\end{eqnarray}
Later on, this will be used to estimate the size of the muon polarization
from $W$-exchange-induced $G_R$ interaction. 

  An effective pseudo-scalar four-Fermi interaction can be 
 induced at one-loop by invoking  a $\tilde{t}_L$--$\tilde{t}_R$ 
insertion (see Fig.~\ref{fig:diagrams}(b)). 
To linear order in the external momenta, the $W$-exchange diagram gives 
\cite{wn}
\begin{eqnarray}
{\cal L}_2 &= & 
           \frac{4G_F}{\sqrt{2}}
           [\frac{C_1}{m_s} ( p_s - p_u)^{\alpha}  
        +  \frac{C_2}{m_{\mu}} ( p_s + p_u)^{\alpha}]
      (\overline{s}_L u_R)(\overline{\nu}_L \gamma_{\alpha}\mu_L) 
         + h.c., \label{eq:W}
\end{eqnarray}
where $p_s$ and $p_u$ are the momenta of the $s$  and $u$ quarks
respectively, and $C_{1,2}$ are given by
\begin{eqnarray}
C_1 & = &  \frac{\alpha_s}{36\pi} I_1  
\frac{m_s m_t(A_t-\mu\cot\beta)}{m_{\tilde{g}}^3} {V^{SKM}_{33}}^* 
{V^{D_L}_{32}}^* V^{U_R}_{31}  \\
C_2 & = & \frac{\alpha_s}{36\pi} I_2 
\frac{m_{\mu} m_t(A_t-\mu\cot\beta)}{m_{\tilde{g}}^3} {V^{SKM}_{33}}^* 
{V^{D_L}_{32}}^* V^{U_R}_{31} ,
\end{eqnarray}
with
\begin{eqnarray}
 I_1 & = & \int_0^1 dz_1 \int_0^{1-z_1} dz_2 \frac{24z_1(1-z_1-z_2)}{[
\frac{m_{\tilde{t}}^2}{m_{\tilde{g}}^2}z_1+
\frac{m_{\tilde{b}}^2}{m_{\tilde{g}}^2}z_2+
(1-z_1-z_2)]^2}  \label{eq:I1} \\
 I_2 & = & \int_0^1 dz_1 \int_0^{1-z_1} dz_2 \frac{24z_1(z_1-z_2)}{[
\frac{m_{\tilde{t}}^2}{m_{\tilde{g}}^2}z_1+
\frac{m_{\tilde{b}}^2}{m_{\tilde{g}}^2}z_2+
(1-z_1-z_2)]^2} \label{eq:I2}.
\end{eqnarray}
Both integrals $I_{1,2}$ are equal to one at 
$\frac{m_{\tilde{t}}}{m_{\tilde{g}}}=\frac{m_{\tilde{b}}}{m_{\tilde{g}}}=1$,
and both increase as $\frac{m_{\tilde{t}}}{m_{\tilde{g}}}$
and/or $\frac{m_{\tilde{b}}}{m_{\tilde{g}}}$ decreases from one.
For example, 
$I_1 \sim 4$ and 
$I_2 \sim 8$ when $\frac{m_{\tilde{t}}}{m_{\tilde{g}}}
=\frac{m_{\tilde{b}}}{m_{\tilde{g}}}=\frac{1}{2}$.
The  functions $I_1$ and $I_2$ are plotted in Fig.~\ref{fig:4I} for the
case $m_{\tilde{t}}=m_{\tilde{b}}$.

 Notice that the $C_2$ term in Eq.~(\ref{eq:W}) 
can be rewritten as an effective scalar
interaction by use of the Dirac equation for the external leptons. 
The term $(p_s - p_u)^{\alpha} (\overline{s}_Lu_R)$ in Eq.~(\ref{eq:W})
 can be Gordon decomposed into a tensor piece,
a left-handed current piece, and a 
 right-handed current piece.  
The tensor piece can be neglected as the tensor form factor is expected
to be  small 
\footnote{See however Ref.~\cite{gengT} for a discussion of tensor effects.}.
The left-handed piece, having the same structure as the standard model
interaction, does not contribute to $P^{\bot}_{\mu}$ 
(see Table~\ref{tab:signatures}).
The relevant operators of the  effective lagrangian 
${\cal L}_2$ can thus be rewritten as
\begin{eqnarray}
 {\cal L}_2 & = & 
 - \frac{4G_F}{\sqrt{2}} C_1 (\overline{s}_R \gamma^{\alpha} u_R)
      (\overline{\nu}_L \gamma_{\alpha}\mu_L)
 - \frac{4G_F}{\sqrt{2}} C_2 
(\overline{s}_L u_R)(\overline{\nu}_L \mu_R)
 + \cdots , \label{eq:L2S}
\end{eqnarray}
where $C_1$ and $C_2$ measure the strengths of the induced $G_R$ and 
$G_P$ interactions respectively.

   The ${\cal L}_2$ contribution to $\Delta_R$ and $\Delta_P$ can be
read off from Eqs.~(\ref{eq:L2S},\ref{eq:delR},\ref{eq:delP}),
\begin{eqnarray} 
\Delta_R|_{{\cal L}_2} 
 & = & - \frac{\alpha_s}{36\pi} I_1 
\frac{m_s m_t(A_t-\mu\cot\beta)}{m_{\tilde{g}}^3} \times
\frac{[{V^{SKM}_{33}}^* {V^{D_L}_{32}}^* V^{U_R}_{31}]}{\sin\theta_c} 
\label{eq:DeltaRL2} \\
\Delta_P|_{{\cal L}_2}  
 & = & - \frac{\alpha_s}{36\pi} I_2  
\frac{m_K}{(m_s+m_u)} \times
\frac{m_K m_t(A_t-\mu\cot\beta)}{m_{\tilde{g}}^3} \times
\frac{[{V^{SKM}_{33}}^* {V^{D_L}_{32}}^* V^{U_R}_{31}]}{\sin\theta_c} .
\label{eq:DeltaPL2}
\end{eqnarray}
It is readily seen that $\Delta_R|_{{\cal L}_2}$ is suppressed relative to 
$\Delta_R|_{{\cal L}_1}$ at least by a factor of $m_b/m_s\sim 30$ and will
not be considered. By taking the sparticle masses to be about 
$100 \; \mbox{GeV}$ and assuming maximal squark family mixings,
the size of $\Delta_P|_{{\cal L}_2}$ is 
found to be at most of order $10^{-4}$, and thus
the contribution to the averaged transverse  muon polarization is at best
of order $10^{-5}$. 
For these reasons, the effective lagrangian ${\cal L}_2$ will not be
considered further.

  We note in passing that a $\tilde{b}_L$--$\tilde{b}_R$ insertion
in a diagram similar to Fig.~\ref{fig:diagrams}(b) 
can also induce effective $G_R$ and $G_P$ interactions analogous in form to
Eqs.~(\ref{eq:L2S})-(\ref{eq:DeltaPL2}).
However, $\Delta_R$ will be suppressed by $\frac{m_u m_b}{m_s m_t}
\sim 10^{-3}$
relative to that of  Eq.~(\ref{eq:DeltaRL2}), whereas the induced $\Delta_P$ 
can at most be comparable to that of Eq.~(\ref{eq:DeltaPL2}) in the large
$\tan \beta$ limit when the left-right mixings in the top and bottom squarks
could be of the same order of magnitude. Their contributions
to the muon polarization are also negligible. 

 In contrast to $\Delta_R|_{{\cal L}_2}$ and $\Delta_P|_{{\cal L}_2}$,
the magnitude of $\Delta_R|_{{\cal L}_1}$ can be enhanced by a large
$\tan \beta$. The present data constrains 
$\tan \beta/m_{H} < 0.52 \; \mbox{GeV}^{-1}$ \cite{GHN},
where $m_H$ is the mass of the charged Higgs boson.  
For $m_H=100\;\mbox{GeV}$, we can take $\tan \beta=50$.
The muon polarization also depends on the squark mixings
$|V^{D_L}_{32}|$ and $|V^{U_R}_{31}|$ (taking $|V^{SKM}_{33}|\sim 1$).
As pointed out earlier, the $V^U$'s are constrained by $D\overline{D}$ mixing
only in the product of $V^{U_{L,R}}_{31}$ and $V^{U_{L,R}}_{32}$, and 
$|V^{U_R}_{31}| ={\cal O}(1)$ is still allowed.
On the other hand, if assuming $|V^{D}_{33}| ={\cal O}(1)$,
the FCNC process $b \rightarrow s \gamma$ can put a bound on
$|V^{D}_{32}|$ from the gluino diagram.  
However, other SUSY contributions to $b \rightarrow s \gamma$, 
 including the charged Higgs and chargino contributions \cite{bsgamma}, 
can dominate  over the 
gluino effect and render the bound on $|V^{D}_{32}|$ meaningless. 
This  is particularly true if the chargino is relatively light.

Recall that the integral $I_0$ can be of order 10 for reasonable
mass ratios of the squarks and gluinos.
To estimate the maximal size of $\Delta_R|_{{\cal L}_1}$, we therefore take
$I_0=10$, $\tan \beta=50$, $A_t=A_b=|\mu|=m_{\tilde{g}}=100 \; \mbox{GeV}$,
and $|V^{U_R}_{31}|=|V^{D_R}_{32}|=\frac{\sqrt{2}}{2}$ for maximal
squark family mixings. Then we have for its magnitude
\begin{eqnarray}
\Delta_R|_{{\cal L}_1} & \le & 0.01 I_0 \le 0.1 \; .
\end{eqnarray} 

  Depending on the soft photon energy cut, it is seen from 
Fig.~\ref{fig:avepol} that averaging over phase space gives
$\overline{\sigma}$ in the $0 - 0.11$ range, 
whereas $\overline{\sigma_V}$ takes
values between $0$ and  $0.17$. 
For an estimate, we choose 
$E_{\gamma}^{cut} =120 \; \mbox{MeV}$, for which 
 $\overline{\sigma} \simeq 0.1$ and $2\overline{\sigma_V} \simeq 0.3$.
The magnitude of the average muon polarization 
 from the $W$-induced effective $G_R$ interaction is then
\begin{eqnarray}
\overline{P^{\bot}_{\mu}}|_{{\cal L}_1} & \simeq & 
   2 \overline{\sigma_V} Im \Delta_R|_{{\cal L}_1}
 < 3 \times 10^{-2}.
\label{eq:limitpolR}
\end{eqnarray}  
This limit scales as 
$\left( \frac{100 \; \rm{GeV}}{M_{SUSY}} \right)^2 
 \left( \frac{\tan \beta}{50} \right)
 \left( \frac{I_0}{10} \right)
 \left( \frac{Im[{V^{SKM}_{33}}^*V^{U_R}_{31}{V^{D_R}_{32}}^*]}{0.5} \right)$, 
where $M_{SUSY}$ is the SUSY breaking scale.

\subsection{$H^+$ Exchange}

  By using the Dirac equation and Lorentz invariance of the 
amplitude, it can be seen that charged Higgs exchange only gives rise to 
$G_P$ but not $G_R$ interactions.
The $m_t$-enhanced effective four-Fermi interaction is obtained
from the diagram that contains a $\tilde{g}$-$\tilde{t}$-$\tilde{b}$ loop
and the $H^- \tilde{t}_R {\tilde{b}_L}^*$ vertex
(see Fig.~\ref{fig:diagrams}(c)).
It is given by \cite{wn} 
\begin{eqnarray}
{\cal L}_H & = & \frac{4G_F}{\sqrt{2}} C_H 
(\overline{s}_L u_R)(\overline{\nu}_L \mu_R)
 + h.c. ,
\label{eq:LH}
\end{eqnarray}
with
\begin{eqnarray}
C_H & = & - \frac{\alpha_s}{3\pi} I_H \tan \beta
 \frac{m_t m_{\mu}}{m_H^2}  
\frac{\mu + A_t\cot\beta}{m_{\tilde{g}}} 
{V^{H}_{33}}^* {V^{D_L}_{32}}^* V^{U_R}_{31}, 
\label{eq:CH}
\end{eqnarray}
where $V^{H}_{ij}$ is the mixing matrix in the charged-Higgs-squark 
coupling $H^+ {\tilde{u_i}_R}^* \tilde{d_j}_L$, 
and where the integral function $I_{H}$ is given by
\begin{eqnarray}
 I_{H} & = & \int_0^1 dz_1 \int_0^{1-z_1} dz_2 \frac{2}{
\frac{m_{\tilde{t}}^2}{m_{\tilde{g}}^2}z_1+
\frac{m_{\tilde{b}}^2}{m_{\tilde{g}}^2}z_2+
(1-z_1-z_2)}, 
\label{eq:IH}
\end{eqnarray}
which is equal to one at $m_{\tilde{t}}=m_{\tilde{b}}=m_{\tilde{g}}$.
The function $I_H$ is plotted in Fig.~\ref{fig:4I} for the case of
$m_{\tilde{t}}=m_{\tilde{b}}$.
It is seen from the figure that  $I_{H}$ increases relatively slowly 
to 2.3 as $m_{\tilde{t}}$ and $m_{\tilde{b}}$ decrease
 to half the gluino mass.
The contribution to $\Delta_P$ from charged Higgs exchange is given as,
 \begin{eqnarray}
\Delta_P|_{{\cal L}_H}   
  & = & - \frac{\alpha_s}{3\pi} I_H \tan \beta
\frac{m_K}{(m_s+m_u)} \frac{m_K m_t}{m_H^2} \times
\frac{\mu + A_t\cot\beta}{m_{\tilde{g}}} \times
\frac{[{V^{H}_{33}}^* {V^{D_L}_{32}}^* V^{U_R}_{31}]}{\sin\theta_c} .
\label{eq:deltaP}
\end{eqnarray}

To estimate the upper limit on $P^{\bot}_{\mu}$ from Higgs boson exchange,
we assume maximal squark mixings with 
$|V^{D_L}_{32}|=|V^{U_R}_{31}|=1/\sqrt{2}$ and 
take $|V^{H}_{33}| \sim 1$,   $m_{H}=100\; \mbox{GeV}$ and $\tan \beta=50$.
Setting $|\mu|=A_t=m_{\tilde{g}}$, $m_s=150\;\mbox{MeV}$, and $I_{H}=1$, 
we find $|\Delta_P|_{{\cal L}_H} \le 0.03$. 
For $E_{\gamma}^{cut}=120 \; \mbox{MeV}$, 
the magnitude of the average muon polarization is then 
\begin{eqnarray}
\overline{P^{\bot}_{\mu}}|_{{\cal L}_H} & \simeq & 
0.1 Im \Delta_P|_{{\cal L}_H} 
  \le  3 \times 10^{-3} \; .
\label{eq:limitpolP}
\end{eqnarray}  
This limit scales as $ \left( \frac{100 \; \rm{GeV}}{m_H} \right)^2
                \frac{\tan \beta}{50} 
          \frac{Im[{V^{H}_{33}}^* {V^{D_L}_{32}}^* V^{U_R}_{31}]}{0.5}$.

 It should be noted that  the diagram involving  the  
$H^- \tilde{t}_L {\tilde{b}_R}^*$ coupling
is proportional to different squark mixing matrices than those appearing
in Eq.~(\ref{eq:CH}), and that it is suppressed by $m_b$.
To be explicit, the induced four-Fermi operator is of the form
\begin{eqnarray}
{\cal L}_{H,m_b} & = & \frac{4G_F}{\sqrt{2}} C_{H,m_b} 
(\overline{s}_R u_L)(\overline{\nu}_L \mu_R)
 + h.c. ,
\label{eq:LHmb}
\end{eqnarray}
with 
\begin{eqnarray}
C_{H,m_b} & = & - \frac{\alpha_s}{3\pi} I_H \tan \beta
 \frac{m_b m_{\mu}}{m_H^2}  
\frac{\mu + A_b\tan\beta}{m_{\tilde{g}}} 
{V^{{\cal H}}_{33}}^* {V^{D_R}_{32}}^* V^{U_L}_{31}, 
\label{eq:CHmb}
\end{eqnarray}
where $V^{{\cal H}}_{ij}$
denotes the mixing matrix in the charged-Higgs-squark 
coupling $H^+ {\tilde{u_i}_L}^* \tilde{d_j}_R$. 

In the large $\tan \beta$ limit however,   the magnitude of $C_{H,m_b}$ 
may be as large as that of $C_H$,
\begin{eqnarray}
\frac{C_{H,m_b}}{C_H} & = &
\frac{m_b(\mu + A_b \tan \beta)}{m_t(\mu + A_t \cot \beta)}
\frac{
{V^{D_R}_{32}}^* V^{U_L}_{31}} 
{{V^{D_L}_{32}}^* V^{U_R}_{31}} 
\simeq \frac{m_b \tan \beta}{m_t} \frac{A_b}{\mu} 
\frac{
{V^{D_R}_{32}}^* V^{U_L}_{31}}
{{V^{D_L}_{32}}^* V^{U_R}_{31}} ,
\end{eqnarray}
where we have taken $V^{{\cal H}}_{33}=V^{H}_{33}={\cal O}(1)$.
Assume $|{V^{D_R}_{32}}^* V^{U_L}_{31}|\simeq
|{V^{D_L}_{32}}^* V^{U_R}_{31}|$ and $A_b\simeq|\mu|$,
then $|C_{H,m_b}|\sim |C_H|$ when $\tan \beta \simeq  m_t/m_b$.
 Within this region of the parameter space, 
the contribution to $P^{\bot}_{\mu}$ 
from the effective interaction of Eq.~(\ref{eq:LHmb})
can be comparable to the $m_t$-enhanced
effect of Eq.~(\ref{eq:limitpolP}).
However, the maximal size of $P^{\bot}_{\mu}$ is not  
expected to be significantly modified.

\subsection{Discussions}

  Several remarks concerning the effective right-handed current 
contribution $\Delta_R|_{{\cal L}_1}$ of Eq.~(\ref{eq:deltaR}) and
the effective pseudoscalar contribution $\Delta_P|_{{\cal L}_H}$ of 
Eq.~(\ref{eq:deltaP}) are in order.
As noted before, an effective $G_R$ operator can be induced by 
$W$ boson exchange only, whereas the $G_P$ interaction can arise from both 
the $W$ and charged Higgs boson exchange.
However, the $W$-induced $G_P$ interaction never gives a large contribution
to $P^{\bot}_{\mu}$ (see Eq.~(\ref{eq:DeltaPL2})).
In contrast, both the $W$-induced $G_R$ operator 
(Eqs.~(\ref{eq:L1})-(\ref{eq:C0})) and the
charged Higgs induced $G_P$ operator (Eqs.~(\ref{eq:LH})-(\ref{eq:CH})) 
are enhanced in the large $\tan \beta$
limit, and large contributions to the muon polarization are found to be
 possible.

  Firstly, the enhancement effects due to squark family mixings
are readily seen from 
Eqs.~(\ref{eq:deltaR}) and (\ref{eq:deltaP}). 
Without squark family mixings, $\Delta_R$ will be suppressed by
 a factor of $\frac{m_sm_u}{m_tm_b} \simeq 10^{-6}$ relative to the 
result of Eq.~(\ref{eq:deltaR}), and would make the effective right-handed
current effect totally uninteresting. Similarly, $\Delta_P$ of 
Eq.~(\ref{eq:deltaP}) would be 
subject to a suppression factor of $\frac{m_sm_{\mu}}{m_tm_{\mu}}
 =\frac{m_s}{m_t} \simeq 10^{-3}$ in the absence of squark flavor mixings.

  Secondly, the different dependence of $\Delta_R|_{{\cal L}_1}$ 
(Eq.~(\ref{eq:deltaR})) and $\Delta_P|_{{\cal L}_H}$ (Eq.~(\ref{eq:deltaP}))
 on the SUSY parameters is to be noted. On one hand, $\Delta_R|_{{\cal L}_1}$
 involves ${V^{D_R}_{32}}^*$ whereas $\Delta_P|_{{\cal L}_H}$ 
is proportional to 
${V^{D_L}_{32}}^*$, and different models of flavor physics could have 
very different predictions for these  two  squark family mixing elements. 
On the other hand, $\Delta_R|_{{\cal L}_1} \propto
1/M^2_{SUSY}$ and $\Delta_P|_{{\cal L}_H} \propto 1/m_H^2$. 
Depending on the the scales of $M_{SUSY}$  and $m_H$, either 
the $W$ exchange  or charged Higgs exchange can give the leading contribution
to $P^{\bot}_{\mu}$.

  If however assuming $|V^{D_R}_{32}|=|V^{D_L}_{32}|$ and $M_{SUSY}=m_H$,
the effective $G_R$ and $G_P$ operators contribute quite differently
to $P^{\bot}_{\mu}$ because of their different chirality structures. 
As can been seen from Eqs.~(\ref{eq:C0}) and (\ref{eq:CH}), 
$G_R \propto m_tm_b$ and $G_P \propto m_t m_{\mu}$, and therefore
$G_R/G_P \sim m_b/m_{\mu} \sim 50$.
 This difference in strength is due to the fact that Yukawa couplings  
are proportional to the fermion masses whereas gauge interactions 
are not subject to light fermion mass suppression.
We recall that the $m_t$ and $m_b$ factors in $G_R$ come from left-right 
mixings in the top and bottom squark propagators.
To estimate the relative contributions to $P^{\bot}_{\mu}$ from 
$G_R$ and $G_P$, we use Eqs.~(\ref{eq:polP}) and (\ref{eq:polR}),
\begin{eqnarray}
\frac{|P^{\bot}_{\mu,R}|}{|P^{\bot}_{\mu,P}|} 
 & = & \frac{2 \sigma_V}{\sigma}
      \frac{(m_s+m_u)m_{\mu}}{m_K^2} \frac{|Im G_R|} {|Im G_P|}
  \simeq  \frac{1}{5} \frac{m_b}{m_{\mu}} \frac{I_0}{10} \simeq 1-10
\label{eq:ratioRP}
\end{eqnarray}
where  the value $\frac{2 \sigma_V}{\sigma}=3$
has been used, $I_0=1-10$ (see Fig.~\ref{fig:4I}), 
and we have assumed the same magnitude for the phases of  
$G_R$ and $G_P$.
This ratio estimate is not sensitive to the
 value of $\tan \beta$, and for 
$I_0=10$ it agrees with our more detailed calculations
of Eqs.~(\ref{eq:limitpolR}) and (\ref{eq:limitpolP}).

  Our third remark concerns the loop suppression factor $\alpha_s/\pi$
and the enhancement effects due to heavy quarks.
 For this purpose, it is interesting to compare the SUSY loop contribution
and the tree level contribution in the three Higgs doublet model (3HDM)
\cite{KLO}.    
 Assuming maximal squark family mixings and neglecting factors associated
with ratios of the Higgs VEVs,
  we have for SUSY charged Higgs exchange
\begin{eqnarray}
\frac{|P^{\bot}_{\mu,P}|}{|P^{\bot}_{3HDM}|}
& \sim &  \frac{\alpha_s}{3\pi} \frac{m_t m_{\mu}}{m_s m_{\mu}}
 \sim 10,
\end{eqnarray}
and for SUSY $W$-exchange
\begin{eqnarray}
\frac{|P^{\bot}_{\mu,R}|}{|P^{\bot}_{3HDM}|}
& \sim &  \frac{1}{5} \frac{\alpha_s}{36\pi}I_0 \frac{m_tm_b}{m_s m_{\mu}}
  \frac{m_H^2}{M^2_{SUSY}}
 \sim (10 - 10^2) \times \frac{m_H^2}{M^2_{SUSY}}
  \;\;\;\;\;\;\; (\mbox{for $I_0=1-10$}),
\end{eqnarray}
where $P^{\bot}_{3HDM}$ is the muon polarization in the 3HDM, and
where the prefactor of $\frac{1}{5}$ comes from Eq.~(\ref{eq:ratioRP}) 
 and it accounts for the difference in the 
hadronic matrix elements  and kinematic factors between the $G_P$ and $G_R$
interactions.
 Note that $\frac{P^{\bot}_{\mu,R}}{P^{\bot}_{3HDM}}
\propto \frac{m_H^2}{M^2_{SUSY}}$, and this ratio can be suppressed if
the SUSY breaking scale is much higher than the charged Higgs mass. 
 Therefore in the presence of large squark generational  mixings,
the heavy quark enhancement effects can overcome the loop suppression
and the SUSY contribution to $P^{\bot}_{\mu}$ could be larger than the
 3HDM tree-level contribution by one or two orders of magnitude. 

 The fourth and final remark concerns possible correlations of 
the muon polarization
 between the $K_{\mu3}$ decay and the $K_{\mu2\gamma}$ decay.

  We first consider charged-Higgs-exchange effects. For both $K_{\mu3}$ and 
$K_{\mu2\gamma}$ decays, the dominant contribution
to the muon polarization 
could come respectively from the scalar and pseudoscalar
components of  the $m_t$-enhanced effective operator of Eq.~(\ref{eq:LH}). 
It is therefore expected that charged Higgs contribution
to the muon polarization is comparable for both decays, as confirmed by
the explicit calculations of 
Eq.~(\ref{eq:limitpolP}) for the $K_{\mu2\gamma}$ decay and of Ref.~\cite{wn}
for the $K_{\mu3}$ decay. 
Since $G_S=G_P$ from Eq.~(\ref{eq:LH}), it is found  that
the average muon polarization in $K_{\mu3}$ decay is about
a factor of 2 bigger than in $K_{\mu2\gamma}$ decay, but with opposite sign
\footnote{Note that when $\tan \beta$ is as large as $m_t/m_b$,
the charged-Higgs-exchange induced operator of Eq.~(\ref{eq:LHmb})
could be as important as the $m_t$-enhanced operator of 
Eq.~(\ref{eq:LH}). This interaction gives $G_S=-G_P$, and contributes
 to the muon polarization in $K_{\mu3}$ and $K_{\mu2\gamma}$
decays with the same sign.}.
 It is interesting to compare our results with the analysis by  
Kobayashi {\it et. al.} \cite{KLO} in the 3HDM. 
Since  the charged Higgs coupling to light quarks is suppressed by
the quark masses, the dominating four-fermion operator arising from tree 
level charged Higgs  exchange in the 3HDM is expected to be of the form 
$(\overline{s}_R u_L)(\overline{\nu}_L \mu_R)$.
Therefore, $G_S=-G_P$ in the 3HDM and  the muon polarization has the same 
sign for $K_{\mu3}$ and $K_{\mu2\gamma}$ decays.
This sign difference is useful for distinguishing between
the 3HDM and the SUSY-induced interaction in Eq.~(\ref{eq:LH}).
  
  If however, the $W$-exchange-induced $G_R$ interaction dominates over 
the charged Higgs-induced $G_P$ (and $G_S$) interaction 
as we have argued on general
grounds,  $P^{\bot}_{\mu}$ in $K_{\mu2\gamma}$ decay can 
be  much larger than $P^{\bot(\pi)}_{\mu}$ 
in $K_{\mu3}$ decay. In this case, different 
squark mixing matrices are involved in these two decays and the relative
sign of the muon polarization can not be predicted. Such a large difference 
in the magnitude of the muon polarization 
between the $K_{\mu2\gamma}$ and $K_{\mu3}$ decays 
is a special feature of  squark-family mixing and does not occur in the 
3HDM.

\section{Conclusion}
\label{sec:concl}

  As discussed before, the transverse muon polarization 
$P^{\bot(\pi)}_{\mu}$ in the
$K^+ \rightarrow \pi^0 \mu^+ \nu$ ($K_{\mu3}$) decay is sensitive to 
an effective scalar interaction,
 whereas $P^{\bot}_{\mu}$ in the 
 $K^+ \rightarrow \mu^+ \nu \gamma$ ($K_{\mu2\gamma}$) decay can receive 
contributions from effective pseudoscalar, vector, and axial-vector 
 four-Fermi interactions. Due to the V-A nature of the standard model 
charged-current,  only pseudoscalar ($G_P$) and right-handed current ($G_R$)
interactions of Eq.~(\ref{eq:deltal}) contribute to
$P^{\bot}_{\mu}$ in the $K_{\mu2\gamma}$ decay. These two decays are 
therefore complimentary in searching for new sources of $CP$ violation. 
 
  Generally speaking, $G_P$ is suppressed by  light fermion masses.
In our SUSY calculation, $G_P$ is proportional to $m_{\mu}$.
 And $G_R$, being induced by $W$ boson exchange only,
is not subject to light fermion mass suppression. 
The relative enhancement factor
 $G_R/G_P \sim  m_b/m_{\mu}$ can cause the dominance of a 
$G_R$ interaction over a $G_P$ interaction
if  the relevant mass scales and phases are of similar magnitude. 
This could turn out to be one advantage for measuring
the muon polarization in $K^+ \rightarrow \mu^+ \nu \gamma$ decay
over that in $K^+ \rightarrow \pi^0 \mu^+ \nu$ decay, 
as $P^{\bot(\pi)}_{\mu}$ in the $K_{\mu3}$ decay is expected to be
comparable in size to the charged Higgs contribution to 
$P^{\bot}_{\mu}$ in the $K_{\mu2\gamma}$ decay.
For example, we take an optimistic value of 
$|V^{U_R}_{31} V^{D_{L,R}}_{32}| = 1/2$ for the 
squark family mixings and assume $\tan \beta =50$ and a mass scale of 
100 GeV for the charged Higgs and the gluino,
the maximal effects for the $K_{\mu2\gamma}$ decay are 
$|P^{\bot}_{\mu}| \le 3\times 10^{-2}$ for the
$W$-induced $G_R$ interaction and 
$|P^{\bot}_{\mu}| \le 3\times 10^{-3}$ for the charged Higgs induced
$G_P$ interaction.
And these two contributions are larger by factors of
$m_tm_b/m_sm_u \sim 10^6$ and $m_t/m_s \sim 10^3$ respectively
than those in the absence of squark family mixings.
Both limits could be accessible to the KEK E246 experiment and the
proposed BNL experiment.
 However, the FSI effect is possibly as large as $\sim 10^{-3}$ 
and needs to be subtracted out. 
 In comparison, the three-Higgs-doublet model gives comparable
contributions to the muon polarization in  
$K_{\mu3}$ and $K_{\mu2\gamma}$ decays \cite{KLO},  and is therefore easily
distinguished from SUSY models with a dominating 
$W$-exchange induced $G_R$ interaction.

  As the $W$-induced $G_R$ interaction 
($\propto {V^{D_R}_{32}}^* V^{U_R}_{31}/M^2_{SUSY}$) 
and the charged-Higgs induced $G_P$ interaction 
($\propto {V^{D_L}_{32}}^* V^{U_R}_{31} /m^2_{H}$) 
involve different squark mixing matrices and different mass parameters,
the possibility of charged-Higgs-exchange  dominance
should not be discarded.  
For charged Higgs exchange, the $m_t$-enhanced effective interaction 
of Eq.~(\ref{eq:LH}) could have the largest effect on $P^{\bot}_{\mu}$, 
 and it also gives $G_S=G_P$. The 
resulting muon polarizations in  $K_{\mu3}$ and 
$K_{\mu2\gamma}$ decays are comparable in size but opposite in sign.
 By contrast, the dominating effect in the three-Higgs-doublet model
gives $G_S=-G_P$, and the muon polarizations are always comparable
 in size and have the same sign for the two decays \cite{KLO}.
Therefore, the three-Higgs-doublet model is again distinguishable
from SUSY models with charged Higgs exchange dominance.
However, when $\tan \beta$ is roughly as large as  $m_t/m_b$, the 
$m_b$-suppressed four-Fermi operator of Eq.~(\ref{eq:LHmb}),
which gives $G_S=-G_P$,
could be as important as the $m_t$-enhanced operator of
Eq.~(\ref{eq:LH}), and the prediction of the  relative sign
between $P^{\bot(\pi)}_{\mu}$ in the $K_{\mu3}$ decay and 
$P^{\bot}_{\mu}$ in the $K_{\mu2\gamma}$ decay
would then be lost in this particular region of the SUSY parameter space. 

   Unlike the $K_{\mu2\gamma}$ decay for which the branching ratio
is about half of a per cent, the branching ratios for
the radiative  $D \rightarrow l \nu \gamma$ ($l=e,\mu$) and 
$B \rightarrow l \nu \gamma$ ($l=e,\mu,\tau$) decays are too tiny 
for measuring the transverse lepton polarization  $P^{\bot}_l$ 
at future $\tau$-charm and $B$ factories.
On the other hand, due to kinematic suppressions of the
interference amplitude and/or experimental technical difficulties
in measuring the electron polarization, 
other radiative pion and kaon leptonic decays
including $\pi^+ \rightarrow \mu^+ \nu \gamma$, 
$\pi^+ \rightarrow e^+ \nu \gamma$, and
$K^+ \rightarrow e^+ \nu \gamma$,
 are not as suitable for studying $T$ violation at present time.
For these reasons, the radiative $K_{\mu2}$ decay is unique as a probe of
new sources of $CP$ violation.
We have considered large squark family-mixings in supersymmetry, and found 
the polarization measurement 
in the $K_{\mu2\gamma}$ decay and charged meson semileptonic decays
at current and future meson factories 
to be very valuable for probing this region of
SUSY parameter space.

\vskip 0.3in
\acknowledgments

  We would like to thank D.~Bryman, K.~Kiers, T.~Numao, especially 
Y.~ Kuno for valuable conversations, and W.J.~Marciano for 
 suggesting the study of radiative kaon decay and for helpful correspondence.
 This work is partially supported by the Natural Sciences and Engineering
Research Council of Canada.

\addtolength{\baselineskip}{-.3\baselineskip}

\setlength{\baselineskip}{\oldbaselineskip}

\newpage

\begin{table}
\caption{
     Effects of different types of interactions on $P^{\bot}_{\mu}$
in the $K^+_{\mu2\gamma}$ decay. 
 The $G_L$ and $G_R$  are effective operators containing 
respectively left-handed and right-handed quark currents. 
The sign and magnitude  of $\delta$ is model dependent, and it 
 also varies from one type of interaction to another.
A $\surd$ denotes a non-zero contribution to 
the muon polarization.}
\label{tab:signatures}

\begin{tabular}{rcccccc} 
   & 
 $\delta_{IB}$   & 
 $\delta_V$  & 
 $\delta_A$  &  
 $P^{\bot}_{IB-V}$ & $P^{\bot}_{IB-A}$ &
 $~~~~P^{\bot}_{\mu}~~~~$    \\ \hline \hline 
 $~~~~G_S$  
        &  0     &  0     &   0    &  0    &  0    &   0    \\ \hline
$G_P$ & $\delta$ &  0     &   0    & $\surd$ &  $\surd$ & $\surd$  \\ \hline
$G_V$ & 0  & $\delta$ & 0 & $\surd$ & 0 & $\surd$ \\ \hline
$G_A$ & $\delta$ & 0 & $\delta$ & $\surd$ & 0 & $\surd$ \\ \hline 
$G_L$ & $\delta$ & $\delta$ & $\delta$ &  0    &  0    &    0   \\ \hline
$G_R$ & $\delta$ & $-\delta$ & $\delta$ &  $\surd$    &  0    &  $\surd$
\end{tabular}

\end{table}

\newpage

\newpage

\newpage

\newpage

\end{document}